# Terahertz reconfigurable multi-functional metamaterials based on 3D printed mortise-tenon structures

Bo Yu, Lesiqi Yin, Peng Wang & Cheng Gong

# Terahertz reconfigurable multi-functional metamaterials based on 3D printed mortise-tenon structures

Bo Yu, Lesiqi Yin, Peng Wang and Cheng Gong

Institute of Modern Optics, Tianjin Key Laboratory of Micro-scale Optical Information Science and Technology, Nankai University, Tianjin, China

**ABSTRACT**
The emergence of metamaterial has provided an unprecedented ability to manipulate electromagnetic waves, especially in the terahertz band where there is a lack of natural response materials. However, most metamaterials are fixed single function due to the fixed structure at the beginning of design. The paper reports a reconfigurable multi-functional terahertz metamaterial with variable structures based on mortise and tenon mechanism. And a hybrid 3D printing method based on FDM and E-jet is proposed to fabricate the metamaterials, which simplifies the processing process, improves the speed, and reduces the cost compared to traditional semiconductor processing methods. Through flexible mortise and tenon connections, the metamaterial can achieve: (1) narrowband transmission and broadband absorption; (2) perfect reflection; (3) narrowband reflection and broadband absorption. Relying on ingenious design and processing, the multi-functional metamaterials are expected to be widely used in fields such as electromagnetic shielding, radar stealth, communication and so on.

## 1. Introduction

Terahertz (THz) wave is an electromagnetic wave with a frequency of 0.1∼10 THz. Its strong penetration can penetrate most non-polar materials, but its quantum energy is very low, making it safe for the human body and biological tissues. It can also be used for next-generation radar and communication. Therefore, the research on electromagnetic functional devices in terahertz band has attracted extensive attention in academia and industry. However, the primary issue that needs to be addressed is the lack of materials in nature that can respond to terahertz waves. Fortunately, the emergence of metamaterials can solve this problem. They can not only respond to terahertz waves, but also provide unprecedented ability for humans to manipulate terahertz waves. Metamaterial (Pendry et al. 1999; Shelby, Smith, and Schultz 2011; Veselago 1968) is a kind of artificially designed sub-wavelength structural material, which has excellent properties that natural materials do not have or cannot surpass due to the designed structure (Li et al. 2023; Li et al. 2023; Li et al. 2023). Terahertz metamaterials have shown significant potential application value in imaging (Huang et al. 2020; Wang et al. 2021; Wu et al. 2022), tuning (Yang and Lin 2020; Yu et al. 2022; Zhang et al. 2020), communication (Palermo et al. 2022; Shui et al. 2020; Yan et al. 2021), and other fields. However, most metamaterials are usually single-function after design and manufacture, which limits their application.

Therefore, researchers have begun to focus on terahertz metamaterials that can achieve multiple functions. For example, metamaterials combined with phase change materials can change the electromagnetic properties by changing the Fermi energy levels, enabling multiple functions such as polarisation conversion and amplitude tuning (Cai et al. 2018; Gao et al. 2020; Liu et al. 2020; Shen et al. 2018). Moreover, metamaterials based on digital coding can accomplish more complex multi-functional switching and are expected to be applied in wireless communication, radar stealth, imaging display, and other fields (Feng et al. 2023; Liu et al. 2016; Liu et al. 2016; Zhang et al. 2018). However, the above-mentioned metamaterials need to introduce external controllable excitation or instrument. Although there are some metamaterials that can achieve multiple functions without additional excitations by depending

---

on their own structural characteristics (Liu et al. 2021; Liu et al. 2022; Wang et al. 2022; Xu et al. 2019). Their design and switching methods are relatively complex. Moreover, most of the preparation methods of terahertz metamaterial are based on semiconductor technology, which is usually complex, time-consuming, and expensive. Toxic and corrosive chemical reagents are also required during the preparation process, which may pollute the environment. At present, there is still a lack of simple and efficient multi-functional metamaterial design methods, as well as large-scale and cheap preparation processes. In order to overcome these problems, Terahertz metamaterial based on 3D printing technology began to attract researchers' attention (Li et al. 2022; Li, Zhang, and Chen 2021; Shen et al. 2021). Here, we proposed a hybrid 3D printing method based on fused deposition modeling (FDM) and electro-hydrodynamic jet printing (E-jet) technology (Li et al. 2022; Tenggara et al. 2017; Zheng et al. 2022; Zou et al. 2019), exhibiting excellent characteristics of simplicity, speed, affordability, and high accuracy.

The mortise and tenon joint mechanism is a structural method in ancient Chinese architecture that combines the complementary features of three-dimensional geometry, similar to LEGO building blocks (Chen, Qiu, and Lu 2016; Emile et al. 2018; Li et al. 2020), which form a stable frame by inserting the concave and convex parts of two components. Nowadays, connection structures similar to the mortise-tenon joints have been applied to some modern electromagnetic devices (Li et al. 2021; Lu et al. 2020; Yuan et al. 2022). Inspired by this mechanism, a design method of terahertz metamaterial is proposed. Through the modular combination of mortise and tenon cross structures, multiple functions such as narrowband transmission, broadband absorption, and perfect reflection can be achieved in the terahertz band, as well as reconfiguration and switching between various functions. Then, the theoretical model based on dielectric loss, impedance matching, Debye relaxation, and frequency selective surface (FSS) is proposed to explain the working mechanism of the multi-band metamaterials with complementary cross structures. Furthermore, a hybrid 3D printing process based on FDM and E-jet is proposed to prepare samples. Finally, an all-fiber terahertz time-domain spectroscopy system is built to validate the effectiveness of the design method and preparation process, and the testing results are consistent with the simulation results.

## 2. Design and functions

In the field of electromagnetic stealth, transmission-absorption metamaterials (Chen et al. 2017; Pan and Zhang 2022; Yang et al. 2023) could realise the functions that electromagnetic waves in the transmission band can smoothly pass through the materials, while incident waves in the absorption band will be absorbed. It is to solve the problem that traditional metamaterial absorber has only the absorption band but no transmission band. Traditional metamaterial absorbers usually consist of sandwich structures (including the metal resonance layer, the dielectric layer, and the metal substrate). The thickness of the metal substrate is much greater than the skin depth of the electromagnetic wave, resulting in the electromagnetic wave being unable to penetrate the metamaterials. Although it can shield the scanning of enemy radar, it may also hinder its own radar detection. To get rid of this dilemma, researchers have used the schemes of phase change materials (Ge et al. 2022; Ren et al. 2020; Zhang et al. 2022) or lumped elements (Chen et al. 2012; Costa and Monorchio 2012; Gu et al. 2010) to construct metamaterial absorbers with a transmission window and achieve transmission-absorption function, but the above schemes may face problems of difficult preparation and high cost. In addition, schemes using phase change materials usually require complex external excitation devices, and schemes using lumped elements mostly operate in low frequency (microwave) bands.

Therefore, there is still a lack of terahertz multi-functional materials that can be easily produced and regulated. In the previous work, we proposed a terahertz multi-functional metamaterial based on nano-imprinting technology, which can simultaneously achieve broadband absorption and narrowband transmission (Li et al. 2023). Further, we propose another terahertz multi-functional metamaterials based on mortise and tenon mechanism, which could realise not only the functions of transmission-absorption, but also the switchable functions of reflection-absorption and prefect reflection, as shown in Figure 1: (I) Narrowband transmission and broadband absorption are achieved by structure A; (II) Perfect reflection is achieved by the combination of structure A + B; (III) Narrowband reflection and broadband absorption are achieved by the combination of structure A + C. The combination of mortise and tenon structures based on three-dimensional geometric complementarity can quickly and easily switch between the above multiple functions. Moreover, the hybrid 3D printing process combining FDM and E-jet printing can greatly simplify the fabrication process, improve speed, and reduce costs. This kind of multi-functional terahertz metamaterials is expected to be applied to large-scale electromagnetic shielding, radar stealth, electronic countermeasures, and other fields.

Figure 2 shows the design principle and functions of the multi-functional metamaterial based on the ancient

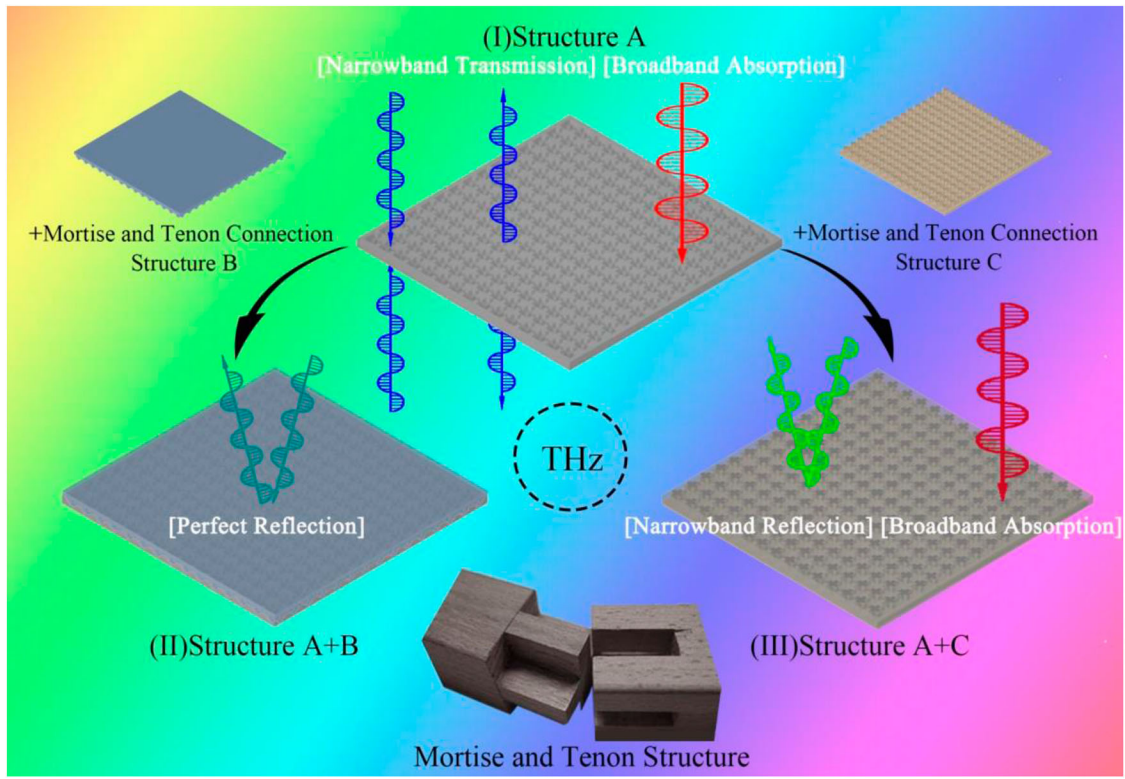

**Figure 1.** Schematic diagram of multi-functional THz metamaterials based on mortise and tenon connection mechanism.

Chinese mortise and tenon joint mechanism. Compared with the simple combination mode of building blocks, the mortise and tenon structure has a more diverse variety of combinations, such as visible tenon, hidden tenon, and penetrated tenon in Figure 2(a). And the cross structure is also a classical structure in the function design of the electromagnetic metamaterial, which could realise the specified function, such as reflection (Chen et al. 2013; Rezaei et al. 2023; Vashistha et al. 2017), transmission (Liu et al. 2019; Xu, Xu, and Lin 2022; Zhu et al. 2012), or absorption (Aydin et al. 2011; Christopher et al. 2023; Kenney et al. 2017). Due to the symmetrical and vertical geometric features, it has the advantages of strong electromagnetic responses, polarisation angle insensitivity, and simple patterns. Therefore, inspired by the similar cross structure, we combined mortise and tenon connection structure and electromagnetic characteristic design of multiple crosses, realising reconfigurable multifunction: narrowband transmission through the cross gap, narrowband reflection through the cross metal, and broadband absorption through the cross dielectric.

Specifically, the polylactic acid (PLA) all-dielectric metamaterial structure A can achieve the transmission-absorption function, as shown in Figure 2(b). The unit of structure A consists of two layers, the upper layer having four 'L' shapes, and the lower layer reminding parts removed a cross structure. The four 'L' structures of four adjacent units can be combined into a single cross structure, so structure A can also be considered as a complementary double cross structure. The unit cycle length $l$ is 2 mm, and the minimum feature size $s$ is 0.5 mm. The cross structure formed by adjacent 'L' shaped structures also maintains a minimum feature size of $s$, and each side has a length of $3 \times s$. The thickness $d$ of the upper and lower layers is also 0.5 mm. Using all-dielectric materials can avoid metal materials from reducing the efficiency of the device, and is conducive to reducing weight (Jahani and Jacob 2016). The upper-layer cross dielectric of structure A could cause the electromagnetic response to realise broadband absorption, while the lower-layer cross gap can act as the FSS to selectively leak the terahertz wave of specific frequencies. In addition, the symmetry cross structure is insensitive to the incident angle, which is significant for the application of electromagnetic stealth (see details in Figure S1 of Supplementary Material).

On the basis of structure A, we combine structure B covered with a silver layer through the hidden tenon connection to achieve a perfect reflection function. Figure 2(c) shows that structure A and structure B are combined into structure A + B through structural complementarity. In a mortise and tenon mechanism, the protruding part is called a tenon and the concave part is called a mortise. The raised cross on the first layer above structure B serves as a tenon, complementing the recessed cross on the first layer above structure A as a mortise. The four 'L' shaped structures on the second layer above structure A serve as a tenon and complement the second layer above structure B as a mortise. The third layer of structure B is a plane PLA with a thickness of 0.5 mm, and the fourth layer is a silver layer with a thickness of 0.1 mm. Therefore, based on the hidden mortise connection structures, structure A and structure B can be combined into structure A + B, as shown in Figure 2(c).

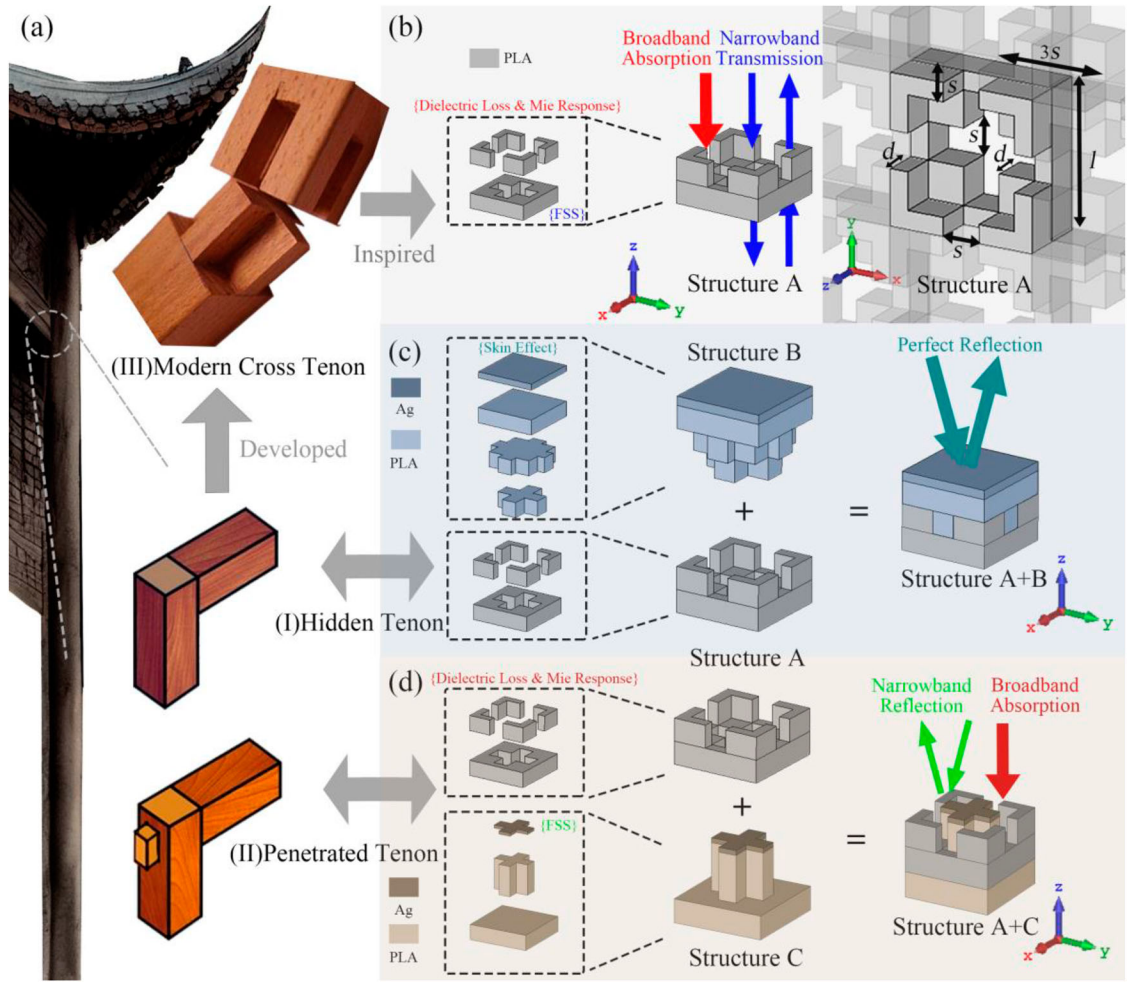

**Figure 2.** Metamaterial structures and combinations: (a) Mortise and tension structure in Chinese ancient building, (b) Structure A; (c) Structure A + B; (d) Structure A + C.

Similarly, by removing structure B, we can combine structure C with structure A through a penetrating tenon connection to achieve narrowband reflection and broadband absorption functions. Structure A acts as a mortise and structure C acts as a tenon. Figure 2(d) describes the process of combining them into structure A + C through mortise and tenon mechanism. Structure C consists of three layers, the first layer being a PLA material base with a thickness of 0.5 mm, the second layer being a PLA cross structure with a thickness of 1.0 mm, and the third layer being a silver metal cross structure with a thickness of 0.1 mm. The cross structure of structure C is smaller than that of structure A, with a single side length of 1.2 mm and a single side width of 0.4 mm. The silver metal cross structure is formed by E-jet of conductive silver paste onto the substrate through a mask pattern, which causes a narrowband reflection peak in the terahertz band to replace the original narrowband transmission peak of structure A. Therefore, based on the penetrated mortise connection structures, structure A and structure C can be combined into a metamaterial structure A + C, as shown in Figure 2(d). Here, structure A improves the bandwidth and efficiency of absorption, and structure C acts as the FSS to selectively reflect the terahertz wave of specific frequencies.

## 3. Simulation and theories

Most metamaterials have complex structures, and it is difficult to obtain their electromagnetic characteristics through analytical methods. Therefore, the general method is to use software to conduct structural modelling and numerical simulation to analyze their electromagnetic characteristics. In this article, we use the multi-physical field simulation software COMSOL Multiphysics, which is numerical simulation software based on the finite element algorithm (FEM). First of all, we model the metamaterial structures, which is a periodic structure with mortise and tenon crosses. Secondly, we set the conductivity of silver; measure the PLA's complex refractive index $n$ in terahertz band; calculate the complex relative dielectric constants $ε_r$, complex relative permeability $μ_r$, and the loss angle $tanδ$ (see the details in Figure S2 of Supplementary Material). And then, the wave equation, initial value, periodic conditions, ports, and scattering boundary conditions are set successively. Next, by simulating the process of linearly polarised terahertz waves incident on metamaterial structure, the complex frequency dependent S parameters can be obtained, including $S_{11}$ representing the reflection coefficient and $S_{21}$ standing for the transmission coefficient. Finally, the electromagnetic response characteristics of metamaterial structure in the terahertz band can be calculated, including the reflection $R$, the transmission $T$, and the absorption $A$, using the following Equation 1:

$$\begin{cases} R = |S_{11}|^2 \\ T = |S_{21}|^2 \\ A = 1 - |S_{11}|^2 - |S_{21}|^2 \end{cases} \quad (1)$$

In the frequency range of 0.1 THz to 0.90 THz, the spectral simulation results of metamaterial structure A are shown in Figure 3(a). As the FSS owing to the lower-layer cross gap, structure A has a transmission peak at a selected low frequency of 0.14 THz, with a peak transmission of 88%. The transmission spectrum of the structure A incident in both directions is consistent. Structure A also maintains broadband absorption at higher frequencies with absorptivity greater than 90% at frequencies above 0.31 THz, which is caused by the upper-layer cross dielectric structure. Therefore, the double-layer-cross metamaterial structure A can achieve the transmission-absorption functions in the terahertz band, that is narrowband transmission at low frequency and broadband absorption at high frequency. It means that when a metamaterial structure A is arranged on the surface of the vehicle, it can absorb terahertz waves emitted by enemy radars, achieving a terahertz broadband stealth effect. The vehicle can transmit and receive radar signals through a specific transmission window in another low frequency band to detect enemy positions or maintain communication with friendly forces.

Figure 3(b) presents the reflection simulation results of structure A + B through a hidden tenon connection. It can be found that the reflection of structure A + B approaches 100%, achieving a perfect reflection effect in the terahertz band. Based on the principle of skin effect, metamaterial structure A + B relies on a silver layer sprayed with E-jet on the surface of structure B to achieve the perfect reflection of terahertz waves, such as for reflecting enemy terahertz interference to protect electronic devices. Skin effect refers to the uneven distribution of current inside a conductor when there is alternating current or alternating electromagnetic field in the conductor, and the current is concentrated in the thin layer on the surface of the conductor. The closer the conductor surface is, the greater the current density is, and the actual current inside the conductor is smaller. This also leads to limited penetration of electromagnetic waves into the conductor, which is called skin depth $Δ$. Its operating band in metamaterials is often much thicker than the geometric scale. According to Equation 2, the penetration depth of 0.10~0.90 THz electromagnetic waves through silver can be calculated $Δ = 6.3 \times 10^{-8} \sim 2.0 \times 10^{-7}$ m, which is far less than the thickness of the silver layer of the structure A + B, enabling perfect reflection. Equation 2 is shown below, where $f$ represents the frequency of the electromagnetic wave, $μ_{Ag} = 4π \times 10^{-7}$ H/m represents the magnetic permeability of silver, $γ_{Ag} = 6.3 \times 10^7$ S/m represents the

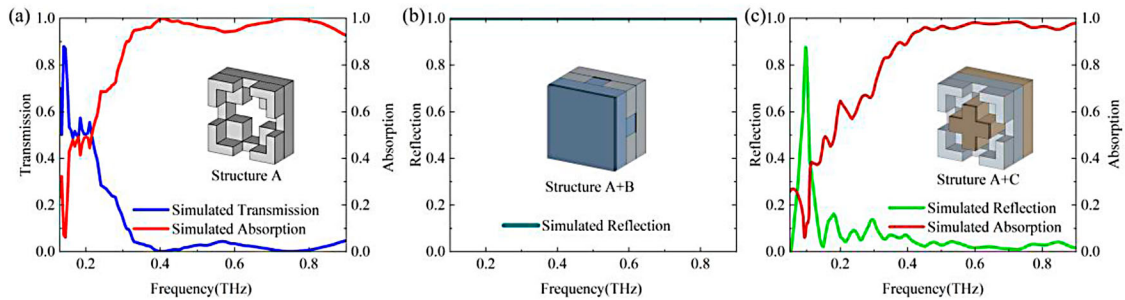

**Figure 3.** Simulation results: (a) Transmission and absorption of structure A; (b) Reflection of structure A + B; (c) Reflection and absorption of structure A + C.

conductivity of silver:

$$\Delta = \sqrt{\frac{2}{2\pi f \mu_{Ag} \gamma_{Ag}}} \quad (2)$$

Figure 3(c) depicts the simulation results of the metamaterial structure A + C in the terahertz frequency range of 0.05~0.90 THz. The structure A + C has a reflection peak at a low frequency of 0.10 THz with a peak reflection of 88%, caused by cross metal acting as the FSS. The structure A + C also maintains broadband absorption at high frequencies with the absorption more than 90% at frequencies at 0.40~0.90 THz, owing to the absorption feature from structure A. Compared with structure A, the connection of penetrated tenon can generate a new function, replacing the narrowband transmission peak at low frequencies with the narrowband reflection peak. Therefore, the metamaterial structure A + C can achieve the reflection-absorption function in the terahertz frequency range, with narrowband reflection at low frequencies and broadband absorption at high frequencies. For instance, when the metamaterial structures A arranged on the surface of a vehicle are switched to structures A + C through modular switching, it can provide its own position to the wingman through a specific reflective window at low frequency. Meanwhile, the vehicle can still absorb terahertz waves emitted by enemy radar, achieving the effect of terahertz broadband stealth.

In order to explain the working mechanism of the multi-band metamaterials with mortise and tenon structures, the theoretical model based on dielectric loss, impedance matching, Debye relaxation, and frequency selective surface is proposed. The broadband absorption effect of multifunctional metamaterial structure can be explained from two aspects: the dielectric loss (see the details in Figure S3 of Supplementary Material) and Mie sub-wavelength resonance (Yang et al. 2023) (see the details in Figure S4 of Supplementary Material). Then, impedance matching theory (Liu et al. 2023) is used to analyze the reduction of reflection and the appearance of transmission peaks at the low frequency of structure A. Extracting the S parameter of structure A from the COMSOL Multiphysics FEM simulation, the complex equivalent impedance $Z$ is calculated by Equation 3:

$$Z = \sqrt{\frac{(1+S_{11})^2 - S_{21}^2}{(1-S_{11})^2 - S_{21}^2}} \quad (3)$$

The real and imaginary parts of the reflection $S_{11}$ are shown in Figure 4(a), and the real and imaginary parts of the transmission coefficient $S_{21}$ are shown in Figure 4(b). Figure 4(c) depicts the complex equivalent impedance $Z$ of structure A. It can be found that the real $Re(Z)$ of the complex equivalent impedance is close to 1 and the imaginary $Im(Z)$ is close to 0 within the frequency range from 0.10 THz to 0.13 THz, which is similar to the complex impedance $Z_0 = 1 + 0j$ in free space. Because the dielectric loss of this band is very low, the incident wave can transmit efficiently through structure A. At 0.13THz, the real and imaginary parts of the complex equivalent impedance Z oscillate violently, and the impedance mismatch causes the transmission to decrease here. Therefore, the transmission band originally located at 0.10∼0.14 THz is split at 0.13 THz, resulting in a sharp transmission peak at 0.14 THz. When the frequency is greater than 0.20 THz, the complex equivalent impedance Z is close to the free space impedance, reflection is suppressed and absorption is enhanced.

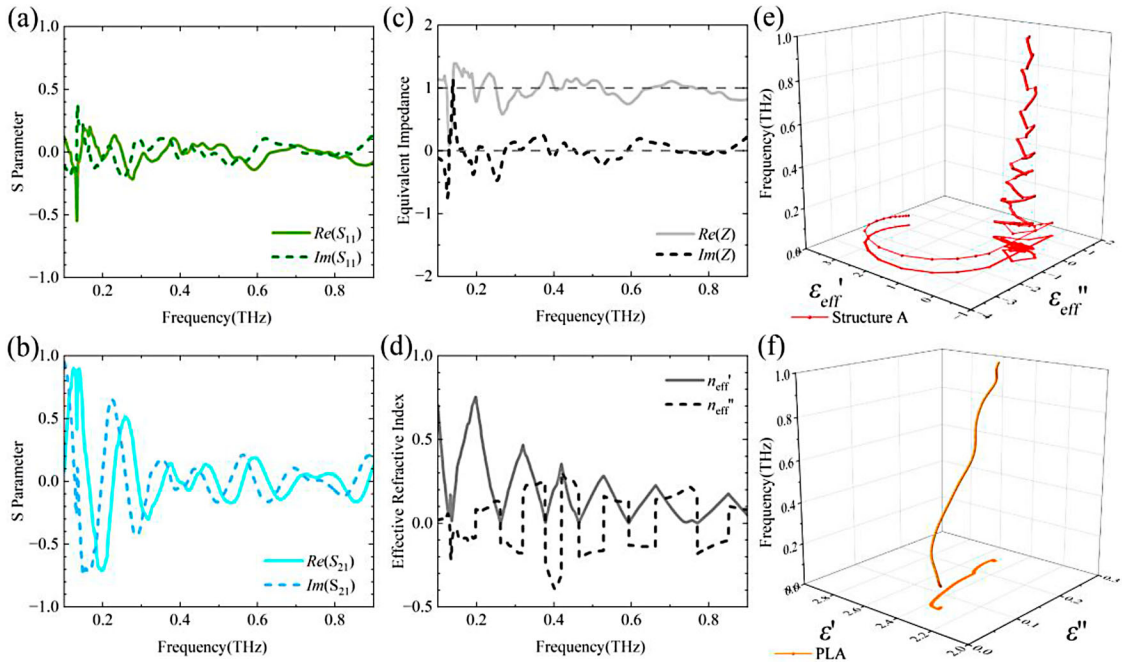

**Figure 4.** Structure A impedance matching calculation: (a) Reflection coefficient; (b) Transmission coefficient; (c) Complex equivalent impedance; (d) Complex equivalent refractive index. Debye relaxation standard: (e)Structure A; (f)Plane PLA.

Secondly, to further explain the broadband absorption mechanism, the complex equivalent refractive index $n_{eff}$ (Smith et al. 2005) of structure A is calculated as shown in Figure 4(d), where $k = 2\pi / \lambda = 2\pi f / c$ representing the vectors of incident terahertz waves, $d$ representing the thickness of each layer of metamaterials in Equation 4:

$$n_{eff} = \frac{1}{k \cdot 2d} \cos^{-1}\left(\frac{1 - S_{11}^2 + S_{21}^2}{2S_{21}}\right) \quad (4)$$

The complex equivalent relative dielectric constant $\varepsilon_{eff} = \varepsilon_{eff}' - i\varepsilon_{eff}''$ of structure A can also be obtained through Equation 5, which can be expressed as the polarisation relaxation standard shown in Figure 4(e) ($\varepsilon_{eff}'$-$\varepsilon_{eff}''$):

$$\varepsilon_{eff} = n_{eff}/Z \quad (5)$$

As a comparison, Figure 4(f) shows the polarisation relaxation criteria for plane PLA ($\varepsilon'$-$\varepsilon''$), which is caused by dielectric loss. According to Debye's theory (Gu et al. 2021) as in Equation 6:

$$(\varepsilon' - \varepsilon_\infty)^2 + (\varepsilon'')^2 = (\varepsilon_s - \varepsilon_\infty)^2 \quad (6)$$

The curve projection staying in the plane of $\varepsilon'$-$\varepsilon''$ will be a bend representing a Debye dipole relaxation. In the upper form, $\varepsilon_s$ is the static dielectric constant, and $\varepsilon_\infty$ is the dielectric constant at an infinite frequency. Figure 4(e) shows more bends than Figure 4(f) indicating a wider electromagnetic absorption due to multiple polarisation relaxation processes due to lots of additional dielectric-air interference interfaces (Zhang et al. 2023). Similarly, due to the multiple gaps between structure A + C and air, terahertz wave is efficiently absorbed by strong Mie resonance between sub-wavelength structures (Cui et al. 2023) (see the details in Figure S5 of Supplementary Material).

Thirdly, the Frequency Selective Surface (FSS) theory plays an important role in the narrowband peak of transmission in structure A or reflection in structure A + C. FSS is a kind of two-dimensional periodic array structure, which is essentially a spatial filter, which is widely used from microwave to visible bands due to its specific frequency selective effect. In the metamaterials with mortise and tenon structure, cross gap and cross metal could be considered as the FSS. The lower-layer cross gap of structure A could act as the FSS to selectively leak the terahertz wave at 0.14THz, while the metal cross of structure A + C could act as the FSS to selectively reflect the terahertz wave at 0.10THz. In summary, we analyze the working mechanism of the multi-band metamaterials through a theoretical model based on impedance matching, Debye relaxation, and FSS theory, where the dielectric loss and Mie resonances realise broadband absorption and the FSS realises narrowband transmission and reflection. The above functions are realised by the combination of various stereo cross structures (cross dielectric, cross gap & cross metal), and they are switched by mortise and tenon structure combination.

## 4. Fabrication and experiments

Compared with the metamaterials fabrication process of photolithography, printed circuit board, mask imprinting, and others, we use the hybrid 3D printing process to construct the mortise and tenon connection structure metamaterials, which has the advantages of simplicity, speed, affordability, and high accuracy. As is shown in Figure 5(a), the FDM device used is an open-source printer (Ultimaker 2 extended+), with a layer resolution of 20 μm, X/Y axis positioning accuracy is 12.5 μm, and Z axis positioning accuracy is 5 μm.

As shown in Figure 5(b), the principle of FDM is through melting the consumables and squeezing them out through gears to form the metamaterial structures in Figure 5(d) (Mohan et al. 2017). The nozzle has a diameter of 0.25 mm, a printing speed of 30 mm/s, and a temperature of 260°C. The opalescent PLA consumables with a diameter of 2.85 mm are used and the flow rate is 0.4 mm/s. Samples of metamaterial structure A, initial structure B, initial structure C, and its visible tenon mask prepared by FDM are shown in Figure 6(a)∼(d), and they are all 15 × 15 units. For example, structure A in Figure 6(a) has a side length of 30 mm, a thickness of 1 mm, and a total printing time of 35 min. Figure 6(f) shows a micrograph of structure A with a 2 mm edge length for each element.

In addition, the print head can be replaced to achieve E-jet printing, as shown in Figure 5(a). The X/Y/Z axis positioning accuracy of the modified E-jet device is 12.5 μm, with a nozzle diameter of 0.17 mm, an applied voltage of 3.5 kV, and a printing speed of 20 mm/s. The consumables used are conductive silver paste with high viscosity and conductivity, and the material flow rate is 3.97 μm/s. The principle is shown in Figure 5(c), when the nozzle is connected to positive high pressure and the base is grounded, an electric field is formed between them. Driven by the electric field, conductive silver paste droplets form a Taylor cone at the nozzle and spray down with a short cone to form a silver layer on the base (Zhao et al. 2016). The specific process is shown in Figure 5(d): (I) A conductive silver paste coating is formed by spraying on the initial structure B, the initial structure C and the mask; (II) Remove the mask on the metamaterial structure C to reveal the silver layer pattern of the cross structure; (III) The conductive silver paste coating is completely cured by heating and drying it with a hot air gun at 50°C. Ultimately, metamaterial structure B and structure C were obtained, which took 25 min in total. Structure C is shown in Figure 6(i) and its micrograph is shown in Figure 6(h). Then, the metamaterial structure A + B is obtained by combining structure B and structure A based on the mortise structure, as shown in Figure 6 (e). The metamaterial structure A + C is obtained by combining structure C and structure A based on the mortise and tenon connection structure, as shown in Figure 6(g).

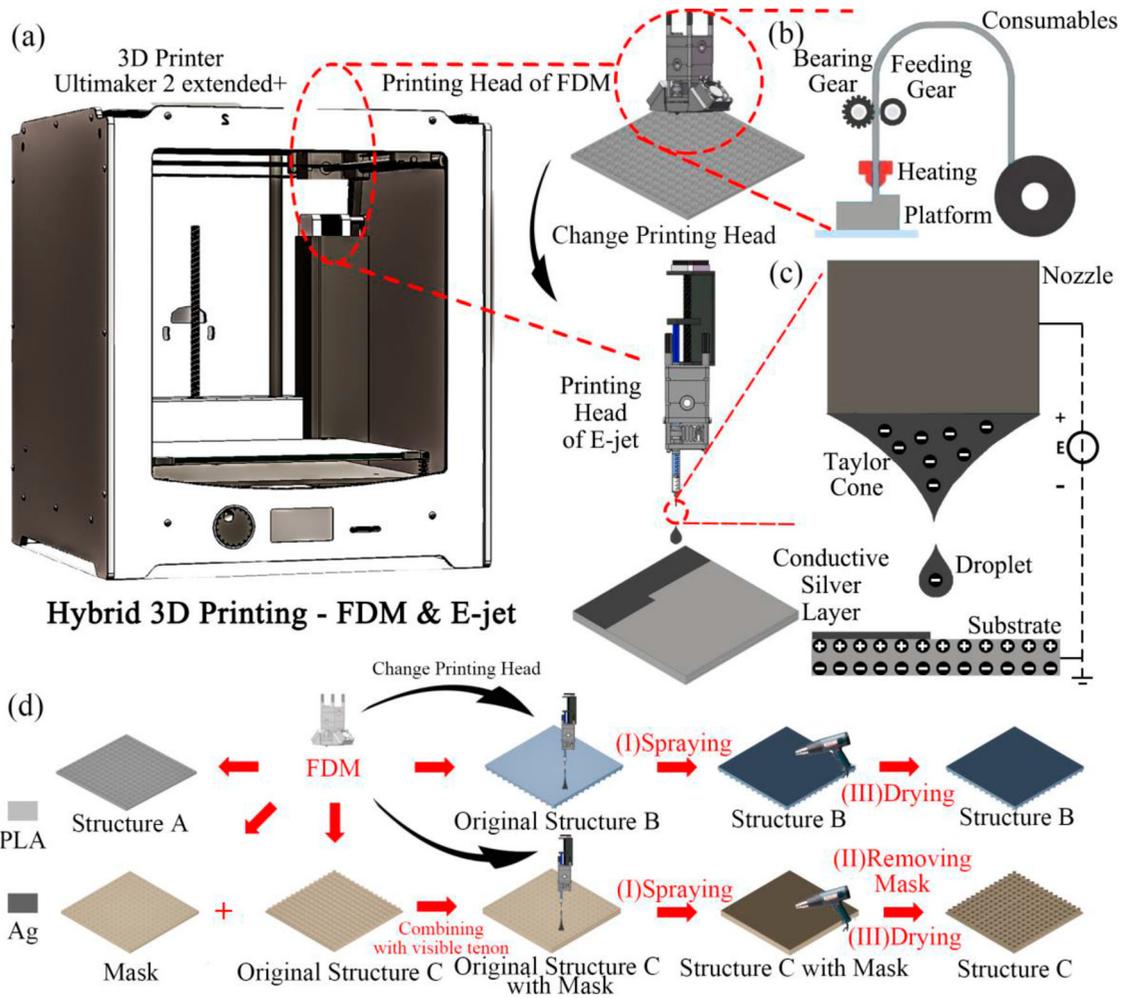

**Figure 5.** Hybrid 3D printing: (a) Device; (b) Schematic of FDM; (c) Schematic of E-jet; (d) Schematic of processes.

The weight of samples can be measured by electronic balance, and we calculate the density of them. Specifically, structure A is 0.55 g and 1.22 g/cm$^3$, structure A + B is 1.33 g and 1.43 g/cm$^3$, and structure A + C is 1.28 g and 1.39 g/cm$^3$. Through mechanical experiments, we measure the various stiffness parameters of metamaterial samples (see the details in Figure S6 of Supplementary Material). Moreover, the Izod impact strength of PLA is 60∼80 J/m, the Rockwell hardness of PLA is 88, and there are fine lines on the surface of the sample caused by the printing head, with a width of approximately 0.006∼0.010 mm and a distance of 0.022∼0.035 mm between them.

Next, a fiber-based terahertz time-domain spectroscopy (THz-TDS) system is established to measure the electromagnetic response spectrum of the metamaterial structures in Figure 7(a). Figure 7(b) shows the diagram of reflection measurement, while Figure 7(c) shows the diagram of transmission measurement. As shown in Figure 7(b)∼(c), antennas $E_1$ and $E_2$ emit terahertz waves, collimate through an off-axis parabolic mirror, and then irradiate the sample. Terahertz waves reflected or transmitted by the sample are focused by an off-axis parabolic mirror and receive terahertz signals from antenna $R_1$ or $R_2$. After further processing of received data, the frequency-domain spectrum of the samples can be obtained (see the details in Figure S5 of Supplementary Material). The reflection $R$ and transmission $T$ could be calculated, while the absorption $S$ could be obtained through Equation 1. To eliminate the interference of moisture in the environment, all processes are conducted at room temperature of 20°C and relative humidity of 0%.

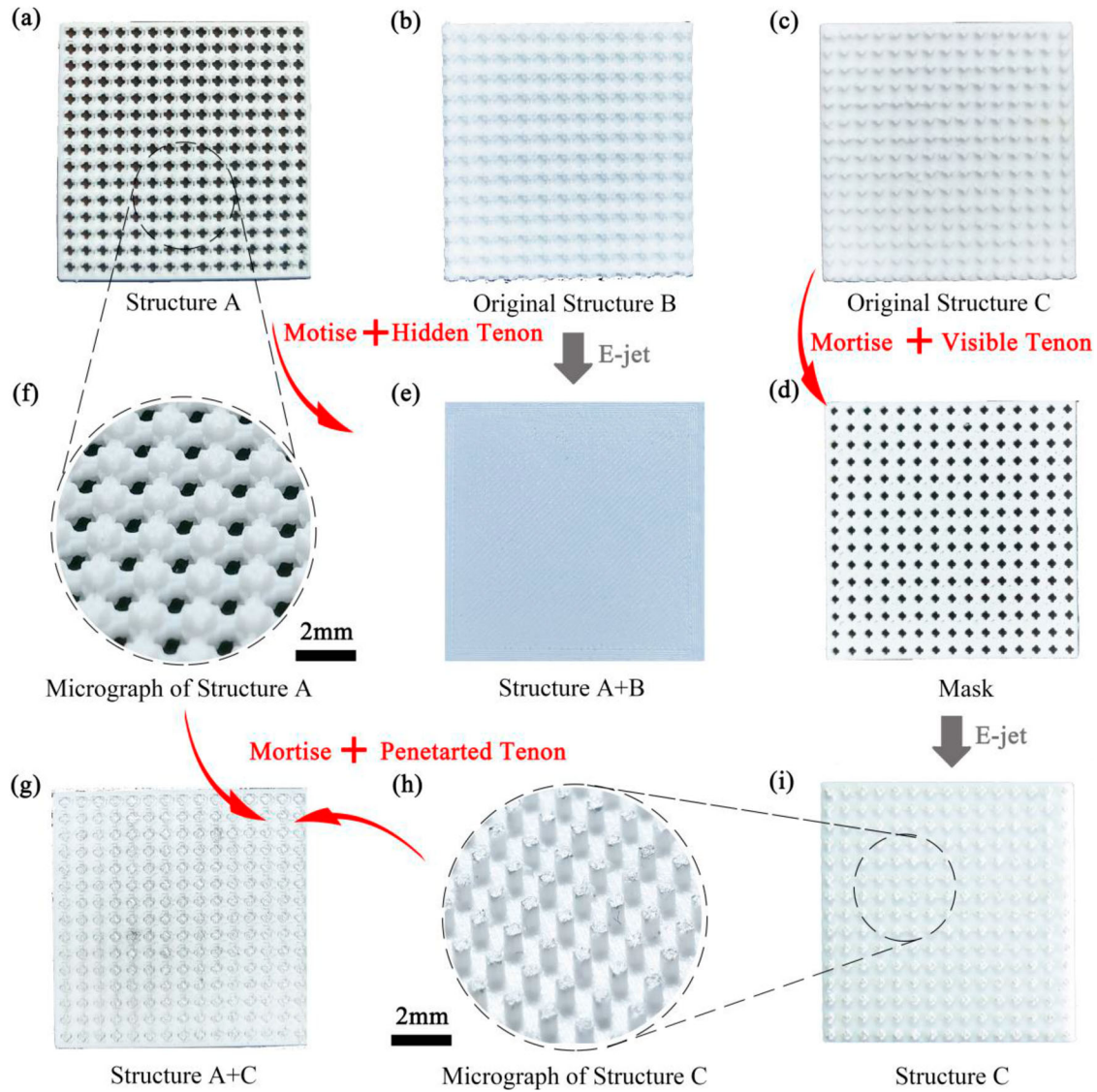

**Figure 6.** Metamaterial samples: (a) Structure A; (b) Original structure B; (c) Original structure C; (d) Mask; (e) Structure A + B; (f) Micrograph of structure A; (g) Structure A + C; (h) Micrograph of structure C; (i) Structure C.

Figure 7(d)∼(f) shows the results of the electromagnetic response characteristics of structure A and structure A + C, respectively, which are in agreement with the results of the FEM simulation. Structure A achieves 86% narrowband transmission at 0.14 THz and more than 90% broadband absorption above 0.28 THz. Structure A + B achieves perfect reflection at 0.20∼0.90 THz. Structure A + C achieves 87% narrowband reflection at 0.10 THz and more than 90% bandwidth absorption above 0.30 THz. However, compared with the simulation results, the measured broadband absorption is stronger, and the transmission peak and reflection peak are slightly different. The main reasons are as follows: (1) The dimension of the metamaterial structure is not completely accurate in the manufacturing process(see the

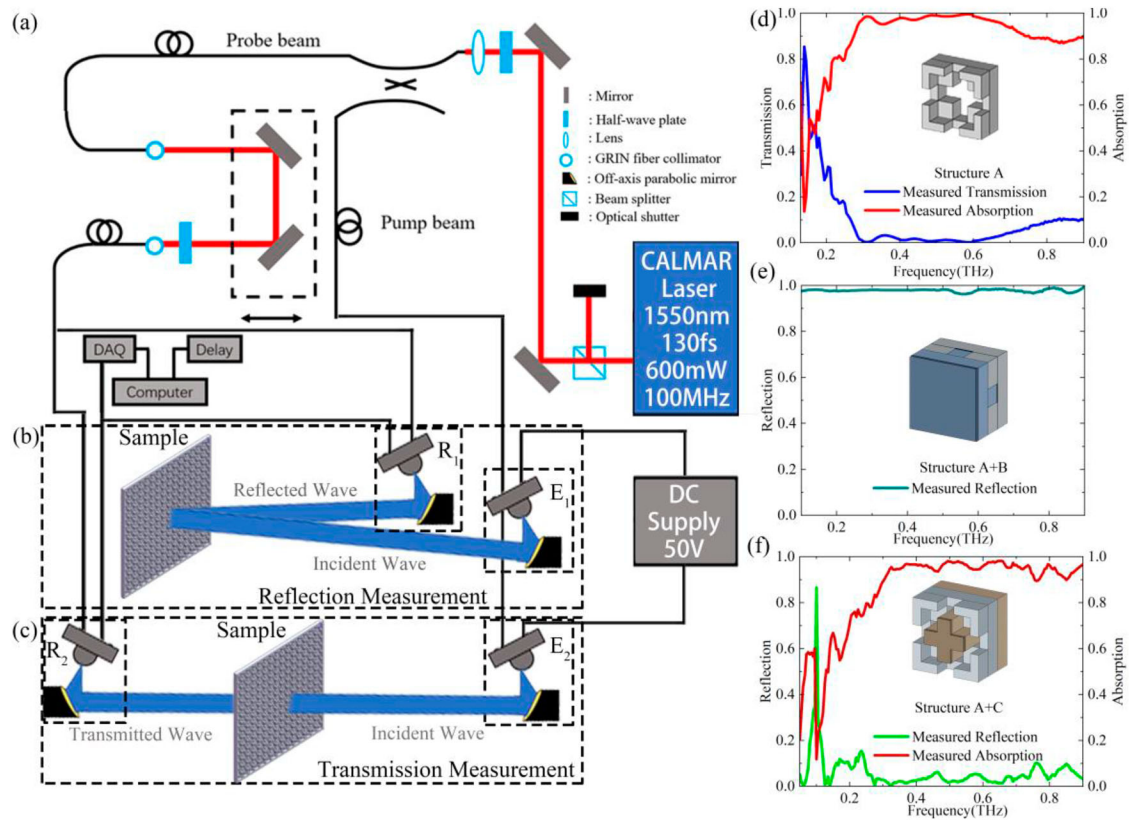

**Figure 7.** THz-TDS: (a) Schematic diagram; (b) Transmission test; (c) Reflection test. Experimental results: (d) Structure A; (e) Structure A + B; (f) Structure A + C.

details in Figure S7 of Supplementary Material); (2) Air gaps exist in the combination of the mortise and tenon structures; (3) A certain deviation of incident angle is required for reflection measurement; (4) Terahertz wave loss is caused by water vapor in the air. Despite these differences, the validity of the mortise and tenon connection structure metamaterials has been demonstrated in the experiments.

In practical applications, this type of multifunctional metamaterial can cover the surface of the vehicle and realise the absorption of the enemy terahertz radar wave, so as to achieve stealth. Besides, it can also cover the surface of our radar or communication equipment. Its broadband absorption function can ensure stealth, while its transmission or reflection function can ensure our normal detection and communication on the battlefield. As shown in Figure 8, based on the modular combination of the mortise and tenon joint mechanism, a variety of functions can be flexibly switched: (I) Narrowband transmission and broadband absorption can be achieved through structure A, which can absorb the detection signal of the enemy terahertz radar, while detection or communication can be carried out through the transmission window at 0.14 THz; (II) Perfect reflection of terahertz band through structure A + B to protect electronic equipment on the vehicle from enemy electromagnetic interference weapons; (III) Narrowband reflection and broadband absorption are achieved through the structure A + C, which absorbs the detection signal of the enemy terahertz radar and at the same time provides the location of the wingman through the reflection window at 0.10 THz. In the ever-changing battlefield, the above functions can be quickly switched to enhance the electromagnetic countermeasures level of the vehicle and improve the survival of personnel and equipment.

## 5. Conclusions

In summary, we demonstrate a modular reconfigurable multi-functional terahertz metamaterial and a hybrid 3D printing process. With the mortise and tenon connection mechanism, the following functions can be achieved and switched: (1) narrowband transmission and broadband absorption, 88% narrowband transmission can be achieved at 0.14 THz, and more than 90% broadband absorption can be achieved at 0.31∼0.90 THz; (2) Perfect reflection at 0.1∼1 THz; (3) narrowband reflection and broadband absorption, 88% narrowband reflection can be achieved at 0.10 THz and more than 90% broadband absorption at 0.40~0.90 THz. To demonstrate its effectiveness, metamaterial samples are prepared using a hybrid 3D printing process based on FDM and E-jet. And an all-fiber terahertz time-domain spectroscopy system is built to test the spectral characteristic, and the testing results are consistent with the simulation results. We believe that the modular reconfigurable multi-functional terahertz Metamaterial and its processing technology can break the limits and serve as a promising platform for advanced functional metamaterials for a wide range of applications, including stealth cloak, radar electronic countermeasures, THz communication, and so on.

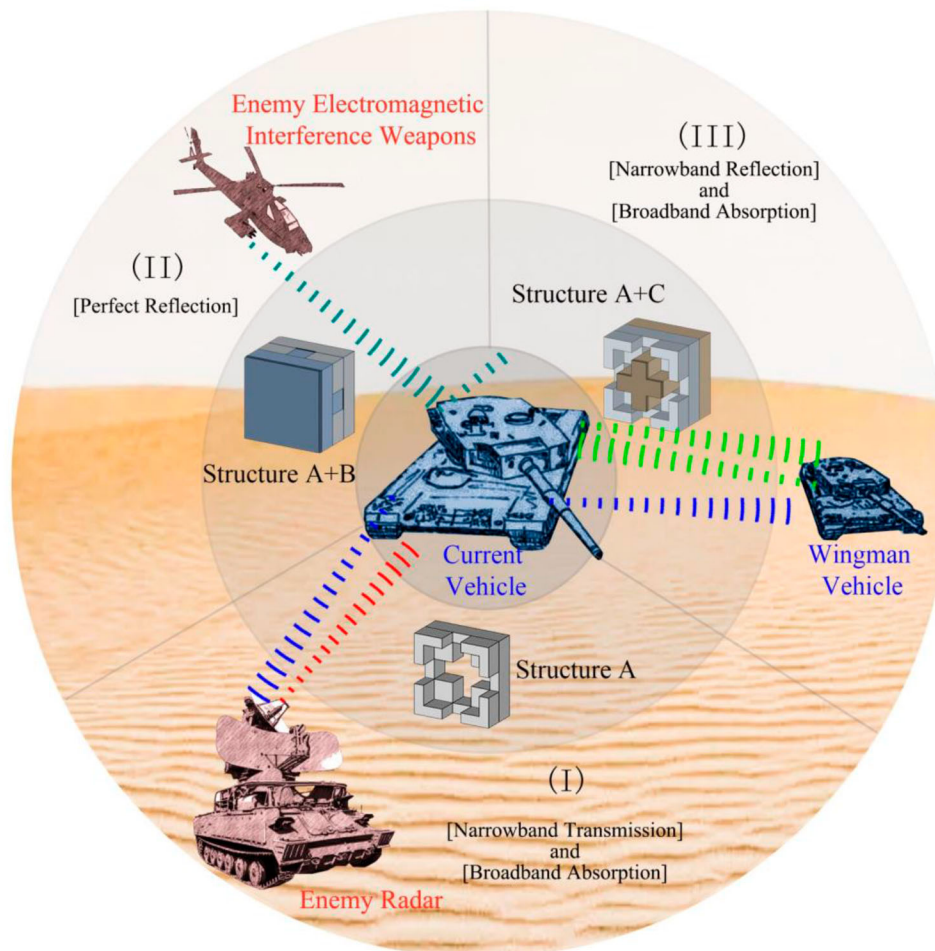

**Figure 8.** Application scenarios of the multi-functional metamaterials.

## Acknowledgment

The authors would like to thank the anonymous reviewers for their feedback and comments.

## Disclosure statement